\documentclass[sigconf]{acmart}

\AtBeginDocument{%
  \providecommand\BibTeX{{%
    \normalfont B\kern-0.5em{\scshape i\kern-0.25em b}\kern-0.8em\TeX}}}

\copyrightyear{2022}
\acmYear{2022}
\setcopyright{acmcopyright}\acmConference[WSDM '22]{Proceedings of the Fifteenth ACM International Conference on Web Search and Data Mining}{February 21--25, 2022}{Tempe, AZ, USA}
\acmBooktitle{Proceedings of the Fifteenth ACM International Conference on Web Search and Data Mining (WSDM '22), February 21--25, 2022, Tempe, AZ, USA}
\acmPrice{15.00}
\acmDOI{10.1145/3488560.3498381}
\acmISBN{978-1-4503-9132-0/22/02}

\usepackage{graphicx}
\usepackage{subfigure}
\usepackage{paralist}
\usepackage{multirow}
\usepackage{upgreek}

\begin{document}
 \fancyhead{}

\title{Multi-Sparse-Domain Collaborative Recommendation via Enhanced Comprehensive Aspect Preference Learning}

\author{Xiaoyun Zhao} 
\affiliation{%
\institution{School of Computer Science}
  \country{Sichuan University, China}
}
\email{zhaoxiaoyun@stu.scu.edu.cn}

\author{Ning Yang}
\authornote{Corresponding Author.}
\affiliation{%
\institution{School of Computer Science}
  \country{Sichuan University, China}
}
\email{yangning@scu.edu.cn}

\author{Philip S. Yu}
\affiliation{%
\institution{Department of Computer Science}
  \country{University of Illinois at Chicago, USA}
}
\email{psyu@uic.edu}



\begin{abstract}
Cross-domain recommendation (CDR) has been attracting increasing attention of researchers for its ability to alleviate the data sparsity problem in recommender systems. However, the existing single-target or dual-target CDR methods often suffer from two drawbacks, the assumption of at least one rich domain and the heavy dependence on domain-invariant preference, which are impractical in real world where sparsity is ubiquitous and might degrade the user preference learning. To overcome these issues, we propose a Multi-Sparse-Domain Collaborative Recommendation (MSDCR) model for multi-target cross-domain recommendation. Unlike traditional CDR methods, MSDCR treats the multiple relevant domains as all sparse and can simultaneously improve the recommendation performance in each domain. We propose a \textit{Multi-Domain Separation Network} (MDSN) and a \textit{Gated Aspect Preference Enhancement} (GAPE) module for MSDCR to enhance a user's domain-specific aspect preferences in a domain by transferring the complementary aspect preferences in other domains, during which the uniqueness of the domain-specific preference can be preserved through the adversarial training offered by MDSN and the complementarity can be adaptively determined by GAPE. Meanwhile, we propose a \textit{Multi-Domain Adaptation Network} (MDAN) for MSDCR to capture a user's domain-invariant aspect preference. With the integration of the enhanced domain-specific aspect preference and the domain-invariant aspect preference, MSDCR can reach a comprehensive understanding of a user's preference in each sparse domain. At last, the extensive experiments conducted on real datasets demonstrate the remarkable superiority of MSDCR over the state-of-the-art single-domain recommendation models and CDR models.
\end{abstract}





\maketitle

\section{Introduction}

\begin{figure}[t]
  \includegraphics[width=\linewidth]{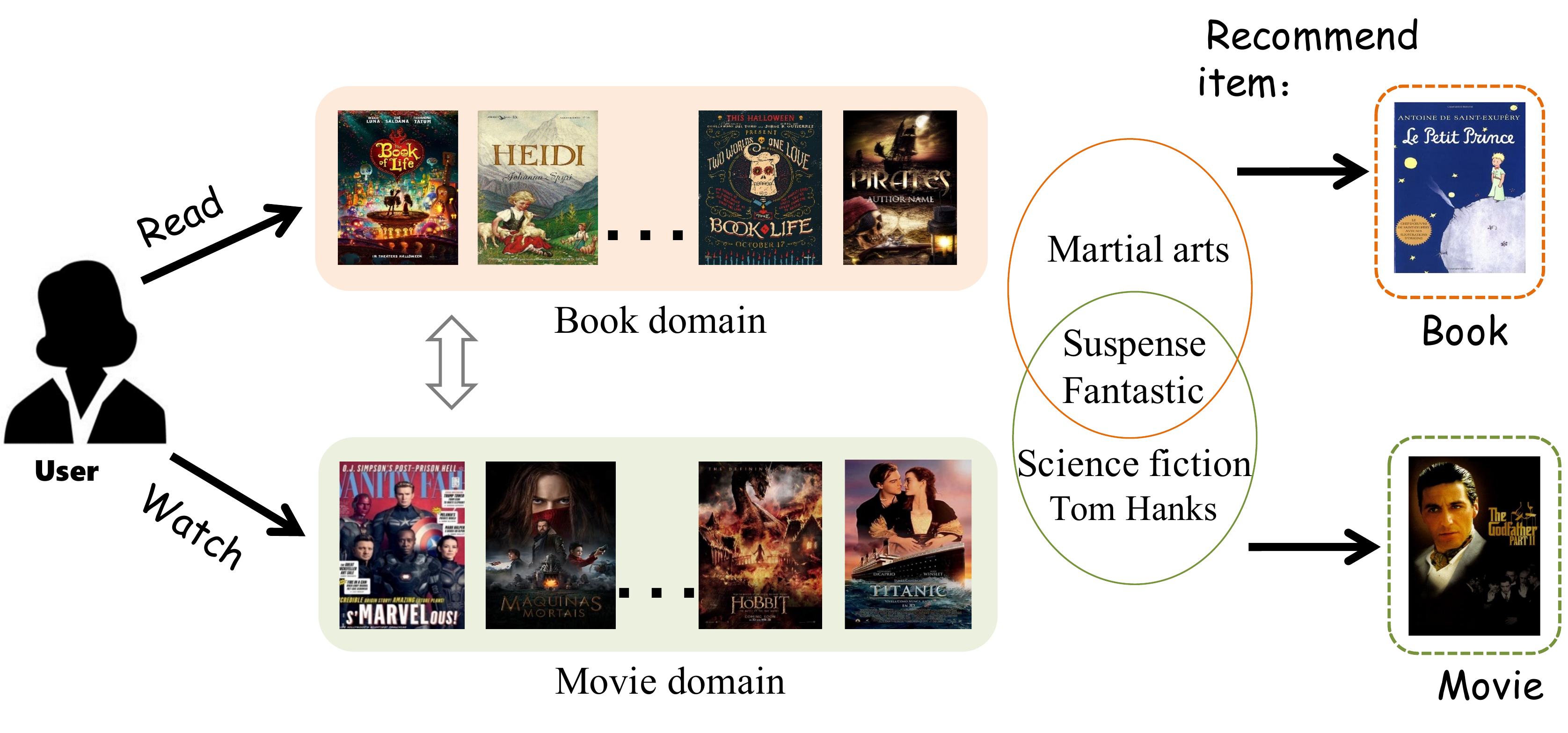}
   \caption{Illustration of Multi-Sparse-Domain Collaborative Recommendation.}
  \label{fig:motivation}
\end{figure}

Cross-domain recommendation (CDR) has been attracting increasing attention of researchers for its ability to alleviate the data sparsity problem in recommender systems \cite{zhu2021cross,wang2019solving}. The common idea of most existing CDR methods is to improve the recommendation performance on a sparse target domain by transferring the information of a relevant source domain with rich data \cite{fu2019deeply,elkahky2015multi,yuan2019darec,zhao2020catn,yu2020semi}, which is called single-target CDR \cite{zhu2019dtcdr}. In real world, however, almost every domain suffers from data sparsity problem due to its ubiquity, which may cause the existing single-target CDR methods to fail in finding a dense auxiliary domain to help the sparse target domain. Recently, dual-target CDR has been proposed to improve recommendation performances on both domains simultaneously, with dual transfer learning of common knowledge (i.e., domain-invariant preference) shared across two domains \cite{singh2008relational,long2012dual,wang2019solving,zhu2019dtcdr,zhao2019cross,zhugraphical2020,li2020ddtcdr}. However, dual-target CDR also assumes that at least one domain has rich information, which still tends to fail under the situations where all domains are sparse.

To tackle this dilemma, we propose a novel recommendation model called \textit{Multi-Sparse-Domain Collaborative Recommendation} (MSDCR) for multi-target CDR. In sharp contrast with traditional CDR methods, MSDCR assumes the multiple relevant domains are \textit{all sparse} and makes them collaborate with each other so that their recommendation performance can be improved simultaneously. Unlike the existing CDR methods which heavily rely on the learning of the domain-invariant preferences, the main idea of MSDCR is to \textit{build a comprehensive understanding of a user's preference to aspects (rather than items) in each domain}, \textit{by capturing not only the user's domain-invariant aspect preferences but also her enhanced domain-specific aspect preferences}. Although the sparse interactions observed in a single domain involves only a small collection of items and can only reveal partial preferences of a user in that domain, we can still \textit{enhance a user's domain-specific preference in a domain by transferring the user's complementary aspect preferences in other domains}. Different from domain-invariant preferences, the complementary preferences refers to those revealed by a user's interactions that are observed only in some domains, but beneficial to the inference of the user's potential interest in other domains where they are not observed. For example, Figure \ref{fig:motivation} shows a user likes science fiction movies, fantasy movies and novels, suspense movies and novels, and martial arts novels. "Fantasy" and "Suspense" are the user's domain-invariant preferences to type (genre) aspect shared across the book domain and movie domain, while "Martial arts" and "Science fiction, Tom Hanks" are the user's aspect preferences specific to the book domain and movie domain, respectively. Although in the book domain we do not observe the user reading science fiction novels, we can reasonably infer her potential interest to them because of her observed preference to science fiction movies in the movie domain. By the same token, we can also guess the user may like martial arts movies based on her interactions with martial arts novels in the book domain. However, learning comprehensive preference for each domain is not easy due to the following challenges:

\begin{compactitem}

\item \textbf{Uniqueness and Commonness} A comprehensive view of a user's preference in each domain is an integration of two ingredients, the domain-specific preference embeddings and the domain-invariant preference embeddings. The domain-specific preference embeddings need to be distinguishable enough to capture the user's unique interest specific to each domain for a meaningful complement between domains. On the contrary, the domain-invariant preference embeddings are supposed to be as indistinguishable as possible, so that the common preferences of a user shared across domains can be revealed for the comprehensive preference learning. 

\item \textbf{Different Complementarity} The complementarity of different aspect preferences is different. On the one hand, the preferences to different aspects in the same domain may have different contributions to the preference learning in another domain. On the other hand, the aspect preferences in other different domains may also have unequal importances to the preference learning for the same domain. 

\end{compactitem}

To address the above challenges, (1) we propose a \textit{Multi-Domain Separation Network} (MDSN) to enforce the \textbf{uniqueness} of the domain-specific aspect preferences, which uses an auxiliary domain separation discriminator to identify the domain label of a domain-specific aspect preference embedding. The objective of MDSN is to make the domain-specific aspect preference embeddings learned from different domains to be as separate from each other as possible, via an adversarial training between the domain-specific aspect preference embedding and the domain separation discriminator. (2) For the enhanced domain-specific aspect preference learning, we propose a \textit{Gated Aspect Preference Enhancement} (GAPE) module to implicitly model the \textbf{complementarity} of a user's latent aspect preferences in different domains. GAPE can generate gate control vectors to regulate the transfer of the user's complementary aspect preferences, by which MSDCR can enhance a user's domain-specific aspect preference in each domain. (3) To obtain a user's comprehensive preference in a domain, we also propose a \textit{Multi-Domain Adaptation Network} (MDAN) for the domain-invariant aspect preference learning, which applies domain adaptation \cite{yu2020semi} to capture a user's preferences \textbf{common} to different domains. (4) Finally, we introduce a multi-task framework to MSDCR, by which MSDCR can be trained under the collaborative supervisions of multiple relevant domains and make multi-target recommendations for a user based on the user's learned comprehensive preference in each domain. The main contributions of this paper can be summarized as follows:

\begin{compactitem}
\item We propose a novel recommendation model called Multi-Sparse-Domain Collaborative Recommendation (MSDCR). Different from traditional CDR methods, MSDCR treats the multiple relevant domains as all sparse and can simultaneously improve the recommendation performance in each domain with comprehensive aspect preference learning which considers both the domain-specific preference and the domain-invariant preference of a user.

\item We propose a \textit{Multi-Domain Separation Network} (MDSN) and a \textit{Gated Aspect Preference Enhancement} (GAPE) module for MSDCR to enhance a user's domain-specific aspect preferences in a domain by transferring the complementary aspect preferences in other domains, during which the uniqueness of the domain-specific preference can be preserved through the adversarial training offered by MDSN and the complementarity can be adaptively determined by GAPE. 

\item We propose a \textit{Multi-Domain Adaptation Network} (MDAN) for MSDCR to capture the domain-invariant aspect preferences, which together with the enhanced domain-specific aspect preferences form a comprehensive view of a user's preference in each domain.

\item We conduct extensive experiments on real datasets from three relevant domains. The experimental results verify the feasibility and the effectiveness of MSDCR.
\end{compactitem}

\begin{figure*}[t]
	\includegraphics[width=0.85\linewidth]{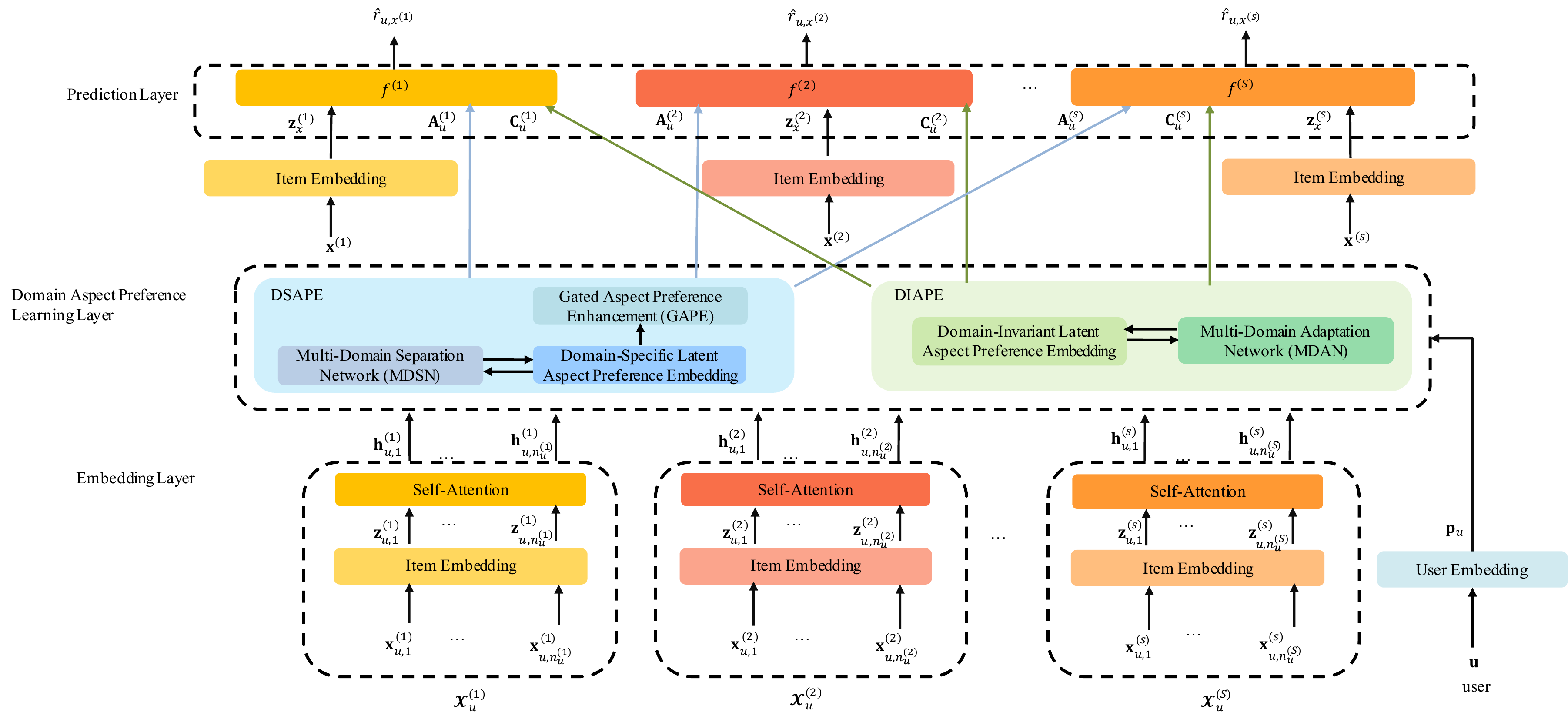}
	\caption{The Architecture of MSDCR.}
	\label{fig:model}
\end{figure*}

\section{Problem Definition}
We consider $S$ relevant domains, and let $\mathcal{D}^{(s)}$ denote the $s$th domain ($1 \le s \le S$). Let $x^{(s)}$ represent an item in $\mathcal{D}^{(s)}$, associated with a raw feature vector $\mathbf{x}^{(s)} \in \mathbb{R}^{d_s}$, where $d_s$ is the dimensionality of the domain-specific features of $\mathcal{D}^{(s)}$. We use a one-hot vector $\mathbf{u} \in \mathbb{R}^{N}$ to represent a user $u$ ($1 \le u \le N$), where $N$ is the number of users shared across domains. Let $\mathcal{X}_u^{(s)}=\{\mathbf{x}_{u,1}^{(s)}, \cdots, \mathbf{x}_{u,n_u^{(s)}}^{(s)}\}$ be user $u$'s collection of $n_u^{(s)}$ interactions with items in domain $\mathcal{D}^{(s)}$, where $\mathbf{x}_{u,i}^{(s)} \in \mathcal{D}^{(s)}$ ($1 \le i \le n_u^{(s)}$) represents $u$'s $i$th interaction (item) in $\mathcal{D}^{(s)}$. Then the target problem of this paper is to learn a function $f^{(s)}$ for each domain $\mathcal{D}^{(s)}$ that can predict the probability $\hat{r}_{u, x^{(s)}}$ of user $u$ will interact with an item $x^{(s)} \in \mathcal{D}^{(s)} \setminus \mathcal{X}_u^{(s)}$, given $u$'s historical interaction collections in $S$ relevant domains $\mathcal{X}_u^{(1)}, \dots, \mathcal{X}_u^{(S)}$.

\section{The Proposed Model}
\subsection{Overview}

Figure \ref{fig:model} shows the architecture of MSDCR, which comprises an embedding layer, a domain aspect preference learning layer, and a prediction layer. At the item embedding layer, MSDCR first transforms the interactions $\{ \mathbf{x}_{u,i}^{(s)}$, $1 \le i \le n_u^{(s)} \}$ in each domain $\mathcal{D}^{(s)}$ ($1 \le s \le S$) to their corresponding latent representations $\{ \mathbf{z}_{u,i}^{(s)} \in \mathbb{R}^d \}$, using a domain-specific item embedding module for $\mathcal{D}^{(s)}$, where $d$ is the embedding dimensionality. To capture the nonlinear relations between the interactions in a domain, MSDCR will further apply a domain-specific self-attention mechanism to $\{ \mathbf{z}_{u,i}^{(s)} \}$ to generate the attentional item embeddings $\{ \mathbf{h}_{u,i}^{(s)} \in \mathbb{R}^d \}$. At the same time, MSDCR will also generate the user's embedding $\mathbf{p}_{u} \in \mathbb{R}^d$ at the embedding layer.

At the domain aspect preference learning layer, with the attentional item embeddings and user embedding as input, MSDCR applies the Domain-Specific Aspect Preference Encoder (DSAPE) to generate a domain-specific latent aspect preference embedding matrix ${\mathbf{A}}_{u}^{(s)} = [ {\mathbf{a}}_{u,m}^{(s)}  ]_{m=1}^{M}$ $ \in \mathbb{R}^{d \times M}$ for each domain, where $M$ is the number of latent aspects and each column vector ${\mathbf{a}}_{u,m}^{(s)} \in \mathbb{R}^d$ represents $u$'s domain-specific preference embedding of $m$th latent aspect of domain $\mathcal{D}^{(s)}$. Each $\mathbf{a}_{u,m}^{(s)}$ represents an enhanced view of the user $u$'s latent aspect preference specific to domain $\mathcal{D}^{(s)}$, which fuses with the complementary information provided by $u$'s aspect preferences in other domains via the GAPE and keeps unique via the MDSN. At the same time, by using the MDAN, the Domain-Invariant Aspect Preference Encoder (DIAPE) will generate a domain-invariant latent aspect preference embedding matrix $\mathbf{C}_{u}^{(s)} = [ \mathbf{c}_{u,m}^{(s)}  ]_{m=1}^{M} \in \mathbb{R}^{d \times M}$ for each domain $\mathcal{D}^{(s)}$, where the column vector $\mathbf{c}_{u,m}^{(s)} \in \mathbb{R}^d$ represents $u$'s domain-invariant preference embedding of $m$th latent aspect in domain $\mathcal{D}^{(s)}$. $\mathbf{A}_{u}^{(s)}$, together with ${\mathbf{C}}_{u}^{(s)}$, forms the comprehensive view of user $u$'s preference in domain $\mathcal{D}^{(s)}$. 

At last, MSDCR will employs a multi-task framework to make the prediction of the probability $\hat{r}_{u, x^{(s)}}$ that $u$ will interact with an item $x^{(s)} \in \mathcal{D}^{(s)} \setminus \mathcal{X}_u^{(s)}$, by feeding the item embedding $\mathbf{x}^{(s)}$, the domain-specific aspect preference embeddings ${\mathbf{A}}_{u}^{(s)}$, and the domain-invariant aspect preference embeddings ${\mathbf{C}}_{u}^{(s)}$ into a Multi-Layer Perceptron (MLP) $f^{(s)}$ which is specific to domain $\mathcal{D}^{(s)} $.

\subsection{Item and User Emebedding}
\subsubsection{Item Embedding}
Given an item represented by original feature vector $ \mathbf{x}_{u,i}^{(s)} \in \mathbb{R}^{d_s}$, we first transform it to an embedding $ \mathbf{z}_{u,i}^{(s)} \in \mathbb{R}^{d}$ with a domain-specific mapping, $\mathbf{z}_{u,i}^{(s)} = \mathbf{W}^{(s)}\mathbf{x}_{u,i}^{(s)}$, 
%\begin{equation}
%\mathbf{z}_{u,i}^{(s)} = \mathbf{W}^{(s)}\mathbf{x}_{u,i}^{(s)},
%\end{equation}
where $\mathbf{W}^{(s)} \in \mathbb{R}^{d_s \times d}$ is the learnable mapping matrix for domain $\mathcal{D}^{(s)}$ ($1\le s \le S$). To capture the nonlinear relationships between the interactions, next we apply a domain-specific self-attention mechanism to the item embeddings to generate attentional item embeddings. To simplify the expression, we horizontally assemble the item embeddings of user $u$ in domain $\mathcal{D}^{(s)}$ into the embedding matrix $\mathbf{Z}_{u}^{(s)} = [ \mathbf{z}_{u,i}^{(s)}  ]_{i=1}^{n_{u}^{(s)}} \in \mathbb{R}^{d \times n_{u}^{(s)}}$, where $\mathbf{z}_{u,i}^{(s)}$ is the $i$th column and $n_{u}^{(s)}$ is the number of items the user interacted with in $\mathcal{D}^{(s)}$. Similarly, let $\mathbf{H}_{u}^{(s)} = [ \mathbf{h}_{u,i}^{(s)}  ]_{i=1}^{n_{u}^{(s)}} \in \mathbb{R}^{d \times n_{u}^{(s)}}$ be the attentional item embedding matrix, where $i$th column $\mathbf{h}_{u,i}^{(s)}$ is the attentional embedding of $ \mathbf{z}_{u,i}^{(s)}$. For each item embedding, we first generate its query vector $ \mathbf{q}_{u,i}^{(s)} \in \mathbb{R}^d$, key vector $ \mathbf{k}_{u,i}^{(s)} \in \mathbb{R}^d$, and value vector $ \mathbf{v}_{u,i}^{(s)} \in \mathbb{R}^d$, with the following transformations: 
\begin{equation}
\mathbf{Q}_{u}^{(s)} = \mathbf{W}_{\text{q}}^{(s)} \mathbf{Z}_{u}^{(s)}, \text{ }
\mathbf{K}_{u}^{(s)} = \mathbf{W}_{\text{k}}^{(s)} \mathbf{Z}_{u}^{(s)}, \text{ }
\mathbf{V}_{u}^{(s)} = \mathbf{W}_{\text{v}}^{(s)} \mathbf{Z}_{u}^{(s)},
\end{equation}
where $\mathbf{Q}_{u}^{(s)} = [ \mathbf{q}_{u,i}^{(s)}  ]_{i=1}^{n_{u}^{(s)}} $, $\mathbf{K}_{u}^{(s)} = [ \mathbf{k}_{u,i}^{(s)}  ]_{i=1}^{n_{u}^{(s)}}$, $\mathbf{V}_{u}^{(s)} = [ \mathbf{v}_{u,i}^{(s)}  ]_{i=1}^{n_{u}^{(s)}} \in \mathbb{R}^{d \times n_{u}^{(s)}}$, and $\mathbf{W}_{\text{q}}^{(s)}$, $\mathbf{W}_{\text{k}}^{(s)}$, $\mathbf{W}_{\text{v}}^{(s)} \in \mathbb{R}^{d \times d}$ are learnable projection matrices. Then $\mathbf{H}_{u}^{(s)}$ can be obtained by 
\begin{equation}
\mathbf{H}_{u}^{(s)} = \text{softmax}\bigg( \frac{\mathbf{Q}_{u}^{(s)} {\mathbf{K}_{u}^{(s)}}^{T}}{\sqrt{d}}\bigg) \mathbf{V}_{u}^{(s)}.
\end{equation}

\subsubsection{User Embedding}
A user $u$'s embedding $\mathbf{p}_{u} \in \mathbb{R}^{d}$ will be obtained with a lookup over a learnable embedding matrix $\mathbf{W}_{\text{u}} \in \mathbb{R}^{d \times N}$, i.e., $\mathbf{p}_{u} = \mathbf{W}_{\text{u}} \mathbf{u}$, 
%\begin{equation}
%\mathbf{p}_{u} = \mathbf{W}_{\text{u}} \mathbf{u},
%\end{equation}
where $N$ is the number of common users and $\mathbf{u} \in \mathbb{R}^{N}$ is a one-hot vector representing user $u$.

\subsection{Domain-Specific Aspect Preference Encoder (DSAPE)}
DSAPE is the crucial component of MSDCR, which generates the enhanced domain-specific aspect preference embedding matrix $\mathbf{A}_u^{(s)}$ for each domain $\mathcal{D}^{(s)}$ with the help of the novel MDSN and GAPE. 

\subsubsection{Domain-Specific Latent Aspect Preference Embedding}
First, based on the attentional item embedding $\mathbf{h}^{(s)}_{u, i}$, we generate the $m$th latent aspect embedding of user $u$'s $i$th interaction item in domain $\mathcal{D}^{(s)}$, $\tilde{\mathbf{e}}_{u,i,m}^{(s)} \in \mathbb{R}^d$, as follow
\begin{equation}
	\tilde{\mathbf{e}}_{u,i,m}^{(s)} = \widetilde{\mathbf{W}}_{\text{a}}^{(m)}(\mathbf{h}_{u,i}^{(s)} \oplus \mathbf{p}_u)+ \tilde{\mathbf{b}}_{\text{a}}^{(m)}, 
\end{equation}
where $\sigma$ is sigmoid function, $\oplus$ represents concatenation, $\widetilde{\mathbf{W}}_{\text{a}}^{(m)} \in \mathbb{R}^{d \times 2d}$ and $\tilde{\mathbf{b}}_{\text{a}}^{(m)} \in \mathbb{R}^{d}$ are learnable parameters. It is noteworthy that the latent aspect embeddings of items in different domains lie in the same latent aspect space, as the projection matrices $\{ \widetilde{\mathbf{W}}_{\text{a}}^{(m)} \}$ and the bias terms $\{ \tilde{\mathbf{b}}_{\text{a}}^{(m)} \}$ ($1 \le m \le M$) are shared across domains.

Intuitively, a user's preference to a latent aspect of domain $\mathcal{D}^{(s)}$ is revealed by the latent aspect embeddings of the $n_u^{(s)}$ items she interacts with in $\mathcal{D}^{(s)}$. Therefore, we can regard the user $u$'s $m$th aspect preference embedding in domain $\mathcal{D}^{(s)}$, $\tilde{\mathbf{a}}^{(s)}_{u, m} \in \mathbb{R}^d$, as a composition of all the $m$th latent aspect embeddings $\{ \tilde{\mathbf{e}}_{u,i,m}^{(s)} \}$ ($1 \le i \le n_u^{(s)}$). In the light of this idea, $\tilde{\mathbf{a}}^{(s)}_{u, m}$ can be generated with the following attention mechanism,
\begin{equation}
\tilde{\mathbf{a}}^{(s)}_{u, m} = \sum_{i=1}^{n_u^{(s)}} \tilde{\alpha}_{u,i,m}^{(s)} \tilde{\mathbf{e}}_{u,i,m}^{(s)}, \ \tilde{\alpha}_{u,i,m}^{(s)} = \frac{ \exp \big( (\tilde{\mathbf{e}}_{u,i,m}^{(s)})^T \tilde{\gamma}^{(m)} \big) } { \sum_{j=1}^{n_u^{(s)}} \exp \big( (\tilde{\mathbf{e}}_{u,j,m}^{(s)})^T \tilde{\gamma}^{(m)} \big) },
\label{Eq:Attention}
\end{equation}
where $\tilde{\gamma}^{(m)} \in \mathbb{R}^d$ is the learnable query vector for $m$th latent aspect. By horizontally assembling the user latent aspect embeddings as columns, we can obtain user $u$'s domain-specific aspect preference embedding matrix $\widetilde{\mathbf{A}}_{u}^{(s)} = [ \tilde{\mathbf{a}}^{(s)}_{u, m} ]_{m=1}^{M} \in \mathbb{R}^{d \times M}$.

\subsubsection{Multi-Domain Separation Network (MDSN)}

In order to avoid the homogenization of a user's domain-specific preferences, which impedes the extraction of complementary information in different domains, the objective of DSAPE is to learn the user's aspect preferences that are unique to each domain. For this purpose, inspired by the idea in \cite{tsai2017adversarial}, we introduce a domain separation discriminator $\psi$ together with DSAPE to form a multi-domain separation network. The domain separation discriminator is implemented as an MLP taking $\widetilde{\mathbf{A}}_{u}^{(s)}$ as input and using a softmax layer to output an $S$-dimensional vector $\hat{\mathbf{y}}_{u,s} = \psi(\tilde{\mathbf{a}}_{u}^{(s)})$ ($\tilde{\mathbf{a}}_{u}^{(s)} = \tilde{\mathbf{a}}^{(s)}_{u, 1} \oplus \cdots \oplus \tilde{\mathbf{a}}^{(s)}_{u, M} $), of which the $i$th component $\hat{y}_{u,s}^{(i)} \in [0,1]$ represents the predicted probability that $\widetilde{\mathbf{A}}_{u}^{(s)}$ comes from domain $\mathcal{D}^{(i)}$. Let $\boldsymbol{\Theta}_{\text{ds}}$ and $\boldsymbol{\Theta}_{\uppsi}$ represent the learnable parameters of DSAPE and the domain separation discriminator, respectively, and the domain label prediction loss can be defined as 
\begin{equation}
 \mathcal{L}_{\text{ds}} (\boldsymbol{\Theta}_{\text{ds}},\boldsymbol{\Theta}_{\uppsi} ) = -\sum_{u = 1}^{N} \sum_{s=1}^{S}\sum_{i=1}^{S} y_{u,s}^{(i)} \log \hat{y}_{u,s}^{(i)},
\end{equation}
where the ground-truth $y_{u,s}^{(i)} = 1$ if $i=s$, otherwise $y_{u,s}^{(i)} = 0$. Then we define the following minimax optimization objective for multi-domain separation:
\begin{equation}
\min_{\boldsymbol{\Theta}_{\text{DS}}}\max_{\boldsymbol{\Theta}_{\uppsi}} \mathcal{L}_{\text{ds}}.
\label{Eq:Lds}
\end{equation}
Note that the optimization objective of the discriminator $\psi$ is to weaken its discriminating ability by maximizing the prediction loss ($\mathcal{L}_{\text{ds}}$). Here the insight is that by adversarially training the DSAPE with the objective of minimizing the prediction loss, the parameters of DSAPE can be adjusted so that the generated domain-specific preference matrices $\{ \widetilde{\mathbf{A}}_{u}^{(s)} \}$ ($1 \le s \le S$) are distinguishable enough even for a poor discriminator $\psi$.

\subsubsection{Gated Aspect Preference Enhancement (GAPE)} 
The domain-specific aspect preference embedding matrices we have learned so far can only uncover partial truth about a user's preference in a domain due to the sparse observations. To overcome this issue, the main novelty of the proposed method is to build an enhanced view of the user's preferences in a domain $\mathcal{D}^{(s)}$, such that we can transfer the complementary aspect preferences in other domains $\{\mathcal{D}^{(s')} \}$ ($s' \ne s$). For this purpose, we first introduce an enhancement gate control matrix $\mathbf{G}_u^{(s , s' )} = [ \mathbf{g}_{u, m}^{(s , s' )} ]_{m=1}^{M} \in \mathbb{R}^{d \times M}$ to regulate the transfer from $\mathcal{D}^{(s')}$ to $\mathcal{D}^{(s)} $, where the $m$th column $\mathbf{g}_{u, m}^{(s , s' )} \in \mathbb{R}^d$ is the gate control vector representing the complementarity of the $m$th latent aspect preference in $\mathcal{D}^{(s')}$ to $\mathcal{D}^{(s)}$. Inspired by the idea in \cite{zhao2020catn}, we compute the gate control matrix as
\begin{equation}
\mathbf{G}_u^{(s , s' )}  = \sigma \bigg( \mathbf{W}_{\text{g}}^{(s')} \big[ ( \widetilde{\mathbf{A}}_{u}^{(s)} \odot \widetilde{\mathbf{A}}_{u}^{(s')}) \oplus ( \widetilde{\mathbf{A}}_{u}^{(s)} - \widetilde{\mathbf{A}}_{u}^{(s')}) \big] + \mathbf{B}_{\text{g}}^{(s')}  \bigg),
%\mathbf{G}_u^{(s , s' )}  = \sigma \big[ \mathbf{W}_{\text{g}}^{(s')} (  \widetilde{\mathbf{A}}_{u}^{(s)} - \widetilde{\mathbf{A}}_{u}^{(s')}) + \mathbf{B}_{\text{g}}^{(s')}  \big],
\end{equation}
where $\mathbf{W}_{\text{g}}^{(s')} \in \mathbb{R}^{d \times 2d}$ and $\mathbf{B}_{\text{g}}^{(s')} \in \mathbb{R}^{d \times M}$ are learnable projection matrix and bias matrix, respectively, and $\odot$ represents element-wise product. 

What does $\mathbf{G}_u^{(s , s' )}$ do for us? Basically, $\mathbf{G}_u^{(s , s' )}$ implicitly models the complementarity of user aspect preferences from two angles, the feasibility and the meaningfulness. At first, we can see that the larger the element-wise product term $\widetilde{\mathbf{A}}_{u}^{(s)} \odot \widetilde{\mathbf{A}}_{u}^{(s')} $ and the difference term $\widetilde{\mathbf{A}}_{u}^{(s)} - \widetilde{\mathbf{A}}_{u}^{(s')} $, the larger the gate control matrix $\mathbf{G}_{u}^{(s , s' )}$ and the more the complementarity of a user $u$'s latent aspect preferences in $\mathcal{D}^{(s')}$ to those in $\mathcal{D}^{(s)}$. The insight here is that a large $\widetilde{\mathbf{A}}_{u}^{(s)} \odot \widetilde{\mathbf{A}}_{u}^{(s')} $ implies both $\tilde{\mathbf{a}}_{u,m}^{(s)}$ and $\tilde{\mathbf{a}}_{u,m}^{(s')}$ are far away from zero vector, i.e., the user $u$'s $m$th latent aspect preferences in $\mathcal{D}^{(s)}$ and $\mathcal{D}^{(s')}$ are both valid (e.g., $u$ cares about the category aspect in both book and movie domains), which makes the complementing to be feasible. At the same time, a large difference $\widetilde{\mathbf{A}}_{u}^{(s)} - \widetilde{\mathbf{A}}_{u}^{(s')} $ means the user preferences to the same aspect in the two domains are different (e.g., $u$ likes category "martial arts" in book domain but category "science fiction" in movie domain), which makes the complementing to be meaningful.

At last, the enhanced aspect preference embedding matrix $\mathbf{A}_u^{(s)} = [ \mathbf{a}^{(s)}_{u, m} ]_{m=1}^{M} \in \mathbb{R}^{d \times M}$ specific to domain $\mathcal{D}^{(s)}$ can be obtained with the following gated fusion of the aspect preferences in other domains:
\begin{equation}
\begin{aligned}
\mathbf{A}_u^{(s)} = \sigma \bigg( \mathbf{W}_{\text{e}}^{(s)} \big[ (\mathbf{G}_u^{(s , 1 )} \odot  \widetilde{\mathbf{A}}_{u}^{(1)} ) \oplus \cdots \oplus (\mathbf{G}_u^{(s , s-1 )} \odot  \widetilde{\mathbf{A}}_{u}^{(s-1)} ) \\
\oplus \widetilde{\mathbf{A}}_{u}^{(s)} \oplus (\mathbf{G}_u^{(s , s+1 )} \odot  \widetilde{\mathbf{A}}_{u}^{(s+1)} ) \oplus \cdots \oplus (\mathbf{G}_u^{(s , S )} \odot  \widetilde{\mathbf{A}}_{u}^{(S)} ) \big] + \mathbf{B}_{\text{e}}^{(s)} \bigg),
\end{aligned}
\end{equation}
where $\mathbf{a}^{(s)}_{u, m} \in \mathbb{R}^d$ is the user $u$'s enhanced $m$th aspect preference embedding in domain $\mathcal{D}^{(s)}$, $\mathbf{W}_{\text{e}}^{(s)} \in \mathbb{R}^{d \times Sd}$ and $\mathbf{B}_{\text{e}}^{(s)} \in \mathbb{R}^{d \times M}$ are learnable projection matrix and bias matrix, respectively.

\newcommand*\xbar[1]{%
  \hbox{%
    \vbox{%
      \hrule height 0.5pt % The actual bar
      \kern0.5ex%         % Distance between bar and symbol
      \hbox{%
        \kern-0.1em%      % Shortening on the left side
        \ensuremath{#1}%
        \kern-0.1em%      % Shortening on the right side
      }%
    }%
  }%
}

\subsection{Domain-Invariant Aspect Preference Encoder (DIAPE)}

The challenge for the domain-invariant preference learning is that a user's common preferences revealed by different domains are similar but still slightly different. For example, a user may like suspense novel and movie with different intension. Therefore it may be inappropriate to generate the common preference by manually summing over the domain-specific ones as does the existing method \cite{liu2020cross}. To address this issue, we propose a Multi-Domain Adaptation Network (MDAN) for DIAPE to learn a domain-invariant preference embedding matrix for each domain.

\subsubsection{Domain-Invariant Latent Aspect Preference Embedding}
The goal of DIAPE is to extract a user's domain-invariant latent aspect preference from each domain. Similar to DSAPE, we first generate the item latent aspect embeddings $\bar{\mathbf{e}}_{u,i,m}^{(s)} \in \mathbb{R}^d$ for domain-invariant preference learning,
\begin{equation}
	\bar{\mathbf{e}}_{u,i,m}^{(s)} = \xbar{\mathbf{W}}_{\text{a}}^{(m)}(\mathbf{h}_{u,i}^{(s)} \oplus \mathbf{p}_u)+ \bar{\mathbf{b}}_{\text{a}}^{(m)}, 
\end{equation}
where $\xbar{\mathbf{W}}_{\text{a}}^{(m)} $$\in \mathbb{R}^{d \times 2d}$ and $ \bar{\mathbf{b}}_{\text{a}}^{(m)}$$\in \mathbb{R}^d$ are learnable parameters, and $1 \le u \le N$, $1 \le i \le n_u^{(s)}$, $1 \le m \le M$. Then we extract the domain-invariant latent aspect preference embedding $\mathbf{c}^{(s)}_{u, m} \in \mathbb{R}^d$ with the similar attention mechanism used in DSAPE (Equation (\ref{Eq:Attention})). At last, by horizontal assembly of the embeddings, we can obtain the domain preference matrix $\mathbf{C}_u^{(s)} = [ \mathbf{c}^{(s)}_{u, m} ]_{m=1}^{M} \in \mathbb{R}^{d \times M}$.  

\subsubsection{Multi-Domain Adaptation Network (MDAN)}
To ensure $\mathbf{C}_u^{(s)}$ to be domain-invariant, we also introduce an auxiliary domain adaptation discriminator $\phi$, which together with the domain-invariant latent aspect preference embedding forms a domain adaptation network. The loss of the domain adaptation network is defined as 
\begin{equation}
\mathcal{L}_{\text{da}}(\boldsymbol{\Theta}_{\text{da}}, \boldsymbol{\Theta}_{\upphi}) = -\sum_{u = 1}^{N} \sum_{s=1}^{S} \sum_{i=1}^{S} z_{u,s}^{(i)} \log \hat{z}_{u,s}^{(i)},
\end{equation}
where $\boldsymbol{\Theta}_{\text{da}}$ and $\boldsymbol{\Theta}_{\upphi}$ are the parameters of DIAPE and the domain adaptation discriminator, respectively, $ \hat{z}_{u,s}^{(i)}$ is the $i$th component of the softmax output $\phi(\mathbf{c}_{u}^{(s)}) \in \mathbb{R}^S$ ($\mathbf{c}_{u}^{(s)} = \mathbf{c}^{(s)}_{u, 1} \oplus \cdots \oplus \mathbf{c}^{(s)}_{u, M} $), and $ z_{u,s}^{(i)}$ is the ground-truth such that $ z_{u,s}^{(i)} = 1$ if $i=s$, otherwise $ z_{u,s}^{(i)} = 0$. Note that the role of the domain adaptation discriminator $\phi$ is to help DIAPE generate indistinguishable preference matrices $\{ \mathbf{C}_u^{(s)}\}$. Therefore, the optimization objective of the multi-domain adaptation network is defined as the following minimax game,
\begin{equation}
\min_{\boldsymbol{\Theta}_{\text{DI}}}\max_{\boldsymbol{\Theta}_{\upphi}} -\mathcal{L}_{\text{da}}.
\label{Eq:Lda}
\end{equation}
It is noteworthy that the adversarial optimization objective of the domain adaptation discriminator is different from that of the domain separation discriminator. In Equation (\ref{Eq:Lda}), we want to train the domain adaptation discriminator to maximize the negative cross entropy (-$\mathcal{L}_{\text{da}}$), which equivalently minimizes the cross entropy to yield a strong discriminator, while in Equation (\ref{Eq:Lds}), we want to maximize the cross entropy ($\mathcal{L}_{\text{ds}}$). Therefore, the adversarial optimization of Equation (\ref{Eq:Lda}) results in a good DIAPE able to generate $\{ \mathbf{C}_u^{(s)}\}$ that are indistinguishable enough to fool a strong domain adaptation discriminator. 

\subsection{Interaction Prediction}

For each domain $\mathcal{D}^{(s)}$, we use an MLP $f^{(s)}$ with a sigmoid function as output to predict the probability that a user $u$ will interact with an item $x^{(s)}$,
\begin{equation}
\hat{r}_{u, x^{(s)}} = f^{(s)} ( \mathbf{x}^{(s)}, \mathbf{a}_u^{(s)}, \mathbf{c}_u^{(s)}; \boldsymbol{\Theta_{\text{f}}^{(s)}}),
\end{equation}
where $\boldsymbol{\Theta_{\text{f}}^{(s)}}$ represents the learnable parameters, $\mathbf{x}^{(s)}$ is the original feature vector of the item, $\mathbf{a}_{u}^{(s)} = \mathbf{a}^{(s)}_{u, 1} \oplus \cdots \oplus \mathbf{a}^{(s)}_{u, M} $ represents $u$'s enhanced domain-specific preference in domain $\mathcal{D}^{(s)}$, and $\mathbf{c}_{u}^{(s)} = \mathbf{c}^{(s)}_{u, 1} \oplus \cdots \oplus \mathbf{c}^{(s)}_{u, M}$ represents $u$'s domain-invariant preference revealed by domain $\mathcal{D}^{(s)}$.

\subsection{Model Training}

For each domain $\mathcal{D}^{(s)}$, we build a training set $\mathcal{T}^{(s)} = \{ (u, x_+, x_-) \}$, where $x_+ \in \mathcal{D}^{(s)}$ and $x_- \in \mathcal{D}^{(s)}$ represent a positive sample and a negative sample of a user $u$, respectively. By applying the popular pair-wise ranking loss BPR \cite{rendle2012bpr}, we define the following objective function for a single domain $\mathcal{D}^{(s)}$,
\begin{equation}
\mathcal{L}_{\text{f}}^{(s)} (\boldsymbol{\Theta_{\text{f}}^{(s)}}) = \frac{1}{|\mathcal{T}^{(s)}|} \sum_{(u, x_+, x_-) \in \mathcal{T}^{(s)}} \log (\hat{r}_{u, x_+} - \hat{r}_{u, x_-}).
\end{equation}
Then the total prediction loss over $S$ domains is
\begin{equation}
\mathcal{L}_{\text{f}} (\boldsymbol{\Theta_{\text{f}}}) = \frac{1}{S} \sum_{s=1}^{S} \mathcal{L}_{\text{f}}^{(s)},
\end{equation} 
where $\boldsymbol{\Theta_{\text{f}}} = \{ \boldsymbol{\Theta_{\text{f}}^{(1)}}, \cdots, \boldsymbol{\Theta_{\text{f}}^{(S)}}\}$. 
At last, by combining $\mathcal{L}_{\text{ds}}$, $\mathcal{L}_{\text{da}}$, and $\mathcal{L}_{\text{f}}$, we can get the following overall optimization objective
\begin{equation}
\min_{\boldsymbol{\Theta_{\text{ds}}}, \boldsymbol{\Theta_{\text{da}}}, \boldsymbol{\Theta_{\text{f}}}} \max_{\boldsymbol{\Theta_{\uppsi}}, \boldsymbol{\Theta_{\upphi}}} \mathcal{L}_{\text{ds}} + \mathcal{L}_{\text{da}} + \mathcal{L}_{\text{f}} + \lambda \lVert \boldsymbol{\Theta}\rVert_2,
\end{equation}
where $\boldsymbol{\Theta}$ represents all trainable parameters and $\lambda$ is a factor to control the contribution of the regularization term. We will apply Adam as the optimizer for the iterative training of MSDCR, where the parameters will be updated alternately. It is noteworthy that MSDCR will be trained under the multi-task framework, so that the training will benefit from the collaboration of the supervision signals from multiple relevant domains. 

\begin{table}[t]
\setlength{\tabcolsep}{0.3mm}
	\begin{tabular}{l | l l l}
	\toprule
		
		Domain                               & Movie          & Book     & Music \\ \hline
		\# Common Users & \multicolumn{3}{c}{800}              \\
		\# Items        & 154,886     & 165,461    & 166,447    \\
		\# Interactions & 93,074      & 29,781     & 30,487     \\
		Sparsity        & 0.075\%     & 0.022\%    & 0.023\%    \\ 
		\bottomrule
	\end{tabular}
\caption{Statistics of datasets}
\label{Tbl:Datasets}	
\end{table}

\begin{table}[t]
	\begin{tabular}{clcl}
		\toprule
		Domain & Feature & Dimensionality              & Type      \\ \hline
		& Director      & $\sim10^3$ & one-hot   \\
		& Writer        & $\sim10^4$ & multi-hot \\
		& Actor         & $\sim10^4$ & multi-hot \\
		Movie  & Type          & 20                          & one-hot   \\
		& Country       & $\sim10^2$ & one-hot   \\
		& Language      & $\sim10^2$ & one-hot   \\
		& Level         & 4                           & one-hot   \\ \hline
		& Writer        & $\sim10^4$ & one-hot   \\
		Book   & Publish       & $\sim10^3$ & one-hot   \\
		& Translator    & $\sim10^3$ & one-hot   \\
		& Level         & 4                           & one-hot   \\ \hline
		& Player        & $\sim10^4$ & one-hot   \\
		& Genre         & $\sim10^2$ & one-hot   \\
		Music  & Album         & $\sim10^4$ & one-hot   \\
		& Media         & 4                           & one-hot   \\
		& Publisher     & $\sim10^3$ & one-hot   \\
		& Level         & 4                           & one-hot   \\ 
		\bottomrule
	\end{tabular}
\caption{Statistics of features of each domain}
\label{Tbl:Features}
\end{table}

\section{Experiments}
\subsection{Experimental Setting}

\subsubsection{Datasets} We collect three datasets corresponding to the Movie, Book, and Music domains of Douban website $\footnote{https://www.douban.com}$. Table \ref{Tbl:Datasets} shows the statistics of the three domains, where sparsity is defined as the ratio of the observed interactions over all possible interactions. By different combinations of the domains, we build three dual-domain scenarios, Movie-Book, Movie-Music, and Book-Music, and one triple-domain scenario Movie-Book-Music. In each scenario, we will check the recommendation performance for the overlapped users in each domain. The features of each domain are summarized in Table \ref{Tbl:Features}. In each scenario, we randomly select 70\% of the data as training set, 10\% as validation set, and the remaining 20\% as testing set. We will repeat such procedure five times and report the average results.

\subsubsection{Baseline Methods}
We compare our MSDCR with eight state-of-the-art recommendation methods, including two single-domain models (NCF, AMCF), two single-target CDR models (MV-DNN, CCCFNet), and four dual-target CDR models (CoNet, DDTCDR, GA-DTCDR, and BiTGCF), which are briefly described as follows:

\begin{itemize}

\item \textbf{NCF} \cite{he2017neural} NCF is a deep learning based model for singe-domain recommendation, which employs an MLP to model nonlinear interactions between user and item embeddings.

\item \textbf{AMCF} \cite{pan2020explainable} AMCF is another single-domain recommendation model, which decomposes item features into multiple aspect embeddings to capture fine-grained user preference.

\item \textbf{MV-DNN} \cite{elkahky2015multi} MV-DNN is a single-target CDR model that can learn domain-invariant user preference by multi-view learning over two domains.

\item \textbf{CCCFNet} \cite{lian2017cccfnet} CCCFNet is a content-based single-target CDR model, which combines matrix factorization and collaborative filtering for domain-invariant representation learning.

\item \textbf{CoNet} \cite{hu2018conet} CoNet can transfer dual knowledge across two domains via the cross connections between them, which improves the recommendation performance on both domains.

\item \textbf{GA-DTCDR} \cite{zhugraphical} GA-DTCDR is a dual-target model that leverages the data of dual domains, transfers the preference of common users across domains, and makes recommendations on both domains.

\item \textbf{DDTCDR} \cite{li2020ddtcdr} DDTCDR is also a dual-target CDR model, which realizes the dual knowledge transfer between two domains, based on a cross-domain preference mapping with orthogonal constraints. 

\item \textbf{BiTGCF} \cite{liu2020cross} BiTGCF is a GCN based dual-target CDR model, which can improve the recommendation performance of both domains simultaneously through a bidirectional knowledge transfer between the two domains.

\end{itemize}

\subsubsection{Metrics and Hyper-parameter Setting}
In this paper, we use Hit Ratio (HR) and Normalized Discounted Cumulative Gain (NDCG) as the metrics, which are widely adopted by the baseline methods. The HR@$k$ is the ratio of the ranking lists where the testing item is ranked in the first $k$ positions, while the NDCG@$k$ further accounts for the position of the hit by assigning higher weight to hits at higher positions. We also adopt the popular \textit{leave-one-out} testing strategy \cite{he2017neural,li2020ddtcdr}, which ranks a positive item of a testing user among her 99 negative items which are randomly sampled.

The hyper-parameters are tuned on validation sets. We set the embedding dimensionality and the number of latent aspects $(d, M)$ as $(128, 5)$, $(32, 7)$, and $(16, 8)$ for scenarios Movie-Music, Movie-Book, and Book-Music, respectively. The MLP used for the prediction function $f^{(s)}$ consists of 2 hidden layers, with 256 and 128 neurons, respectively. The MLPs used for the domain separation discriminator and the domain adaptation discriminator both have 1 hidden layer with 128 neurons. For fairness, the hyper-parameters of the baseline methods are set to their optimal configuration tuned on validation sets.

\begin{table}[t]
	\setlength{\tabcolsep}{0.1mm}
	\begin{tabular}{l|cc|cc|cc|cccc}
		\toprule
	Metric      & \multicolumn{2}{c|}{NDCG@5}        & \multicolumn{2}{c|}{NDCG@10}   & \multicolumn{2}{c|}{HR@5}   & \multicolumn{2}{c}{HR@10}  \\ \hline
	Domain      & Movie           & Music           & Movie           & Music           & Movie           & Music           & Movie           & Music                     \\ \hline
	
	NCF         & 0.1501          & 0.1172          & 0.2230          & 0.1708                & 0.2457          & 0.1958          & 0.3727         & 0.3620            \\
	AMCF        & 0.1719          & 0.1285          & 0.2143          & 0.1958         & 0.2540          & 0.2146          & 0.3862          & 0.4250           \\ \hline
	MV-DNN      & 0.1105          & 0.1265          & 0.1570          & 0.1689          & 0.2015          & 0.2264          & 0.2893          & 0.2769               \\
	CCCFNet     & 0.1469          & 0.1363          & 0.1778          & 0.2058          & 0.2103          & 0.2393          & 0.3169          & 0.2707            \\ \hline
	CoNet       & 0.0645          & 0.1258          & 0.1192          & 0.1558          & 0.1251          & 0.2360          & 0.2585          & 0.2656          \\
	GA-DTCDR    & 0.0378          & 0.0418          & 0.0602          & 0.0596          & 0.0687          & 0.0734          & 0.1217          & 0.1132            \\
	DDTCDR      & 0.1937          & 0.1871          & 0.2216          & 0.2308          & 0.2889          & 0.2405          & 0.3504          & 0.3019             \\
	BiTGCF      & 0.1542          & 0.1361          & 0.1997          & 0.2006          & 0.2671          & 0.2333          & 0.3334          & 0.3152                 \\ \hline
	
	MSDCR        & \textbf{0.2541} & \textbf{0.2258} & \textbf{0.2815} & \textbf{0.3027} & \textbf{0.3105} & \textbf{0.3547} & \textbf{0.4456}          & \textbf{0.4596}    \\ 
	\bottomrule
\end{tabular}
	\caption{Performance comparison in Movie-Music}
    \label{Tbl:mov-msc}
\end{table}

\subsection{Performance Comparison}

Tables \ref{Tbl:mov-msc}, \ref{Tbl:mov-book}, and \ref{Tbl:book-msc} show the performances of dual-target recommendation over the three cross-domain scenarios, Movie-Book, Movie-Music, and Book-Music, respectively. For the single-target CDR methods MV-DNN and CCCFNet, the two domains in each scenario will be the target domain by turn. 

First, we can see that in each scenario, MSDCR significantly improves the recommendation performances on both domains. This result verifies the advantage of MSDCR offered by the gated enhancement of domain-specific aspect preference and the comprehensive aspect preference learning. Second, it is also noteworthy that in most cases the single-domain baseline methods (NCF and AMCF) interestingly perform better than the CDR baseline methods. For example, in each scenario, the performances of the single-target CDR methods on the target domain (i.e., the movie domain) are worse than the sing-domain methods. We argue that this is because in the situations where both domains are sparse, the traditional CDR methods cannot accurately capture user preference in a single domain because of the negative transfer between two sparse domains.  At last, we can also note that DDTCDR and BiTGCF perform better than the other dual-target CDR methods CoNet and GA-DTCDR, partly due to their ability to simultaneously capture domain-specific features and cross-domain features, which is similar to MSDCR. However, different from MSDCR, DDTCDR and BiTGCF still require at least one rich domain, which results in their inferior learning of user preference in multiple sparse domains.

\begin{table}[t]
	\setlength{\tabcolsep}{0.1mm}
	\begin{tabular}{l|cc|cc|cc|cccc}
		\toprule
	Metric      & \multicolumn{2}{c|}{NDCG@5}        & \multicolumn{2}{c|}{NDCG@10}   & \multicolumn{2}{c|}{HR@5}   & \multicolumn{2}{c}{HR@10}  \\ \hline
	Domain      & Movie           & Book           & Movie           & Book           & Movie           & Book           & Movie           & Book                     \\ \hline

NCF         & 0.1515          & 0.1647          & 0.2077          & 0.1882         & 0.2543          & 0.2676          & 0.4424          & 0.4631           \\
AMCF      & 0.2233          & 0.1774          & 0.2106          & 0.2154         & 0.2179          & 0.2670          & 0.4336          & 0.4482           \\ \hline

MV-DNN     & 0.1309       & 0.1269          & 0.1958          & 0.1772         & 0.1833          & 0.1732          & 0.3043          & 0.3009           \\
CCCFNet    & 0.1220      & 0.1335          & 0.1705          & 0.1827         & 0.2012          & 0.1991         & 0.2909           & 0.2986       \\ \hline
CoNet         & 0.0753       & 0.1133          & 0.1252          & 0.1564         & 0.1472          & 0.1993         & 0.2689           & 0.3020        \\
GA-DTCDR    & 0.0398    & 0.0407         & 0.0556          & 0.0787         & 0.0666          & 0.0803         & 0.1015          & 0.1104            \\
DDTCDR      & 0.1906      & 0.1713         & 0.2408          & 0.2213         & 0.2478          & 0.2802         & 0.3603          & 0.3446         \\
BiTGCF      & 0.0890         & 0.1012         & 0.1357         & 0.1508         & 0.1396          & 0.1756         & 0.3324          & 0.2872           \\ \hline

MSDCR        & \textbf{0.2497}          & \textbf{0.2086} & \textbf{0.3065} & \textbf{0.2304} & \textbf{0.2893}          & \textbf{0.3051}  & \textbf{0.5035} & \textbf{0.4371}          \\ 
\bottomrule
\end{tabular}
	\caption{Performance comparison in Movie-Book}
	\label{Tbl:mov-book}
\end{table}

\begin{table}[t]
	\setlength{\tabcolsep}{0.1mm}
	\begin{tabular}{l|cc|cc|cc|cccc}
		\toprule
	Metric      & \multicolumn{2}{c|}{NDCG@5}        & \multicolumn{2}{c|}{NDCG@10}   & \multicolumn{2}{c|}{HR@5}   & \multicolumn{2}{c}{HR@10}  \\ \hline
	Domain      & Book           & Music           & Book           & Music           & Book           & Music           & Book           & Music         \\ \hline
	
	NCF         & 0.1068          & 0.1083          & 0.1602          & 0.2818          & 0.2381          & 0.2387          & 0.4006          & 0.3641  \\
	AMCF        & 0.1347          & 0.1246          & 0.1727          & 0.2656           & 0.2244          & 0.2140          & 0.3425       & 0.3415   \\ \hline
	MV-DNN    & 0.1025        & 0.1257          & 0.1631          & 0.1803          & 0.1985           & 0.2149        & 0.3074          & 0.2953            \\
	CCCFNet   & 0.1153        & 0.1496         & 0.1596           & 0.1703          & 0.2185           & 0.2046        & 0.2977          & 0.3034          \\ \hline
	
	CoNet       & 0.0657         & 0.1152         & 0.1408          & 0.1569           & 0.1663          & 0.2107         & 0.2009         & 0.2563          \\
	GA-DTCDR    & 0.0378    & 0.0362        & 0.0509         & 0.0548            & 0.0621         & 0.0603          & 0.1001         & 0.1207          \\
	DDTCDR      & 0.1538     & 0.1781         & 0.1883         & 0.2049           & 0.2608          & 0.2777          & 0.3912        & 0.3796          \\
	BiTGCF      & 0.1380       & 0.1408         & 0.1812         & 0.2005           & 0.2462          & 0.2702         & 0.4003         & 0.3809          \\ \hline
	
	MSDCR        & \textbf{0.1996} & \textbf{0.2162} & \textbf{0.2325} & \textbf{0.2581} & \textbf{0.3172} & \textbf{0.3114} & \textbf{0.4339}          & \textbf{0.4402}           \\ \bottomrule
\end{tabular}
	\caption{Performance comparison in Book-Music}
	\label{Tbl:book-msc}
\end{table}

\begin{table*}[t]
	\setlength{\tabcolsep}{0.2mm}
\begin{tabular}{l |ccc|ccc|ccc|ccc}
	\toprule
	
	Metric       & \multicolumn{3}{c|}{NDCG@5}                         & \multicolumn{3}{c|}{NDCG@10}                        & \multicolumn{3}{c|}{HR@5}                    & \multicolumn{3}{c}{HR@10}                               \\ \hline
	Domain       & Movie           & Book            & Music           & Movie           & Book            & Music           & Movie           & Book            & Music           & Movie           & Book            & Music                    \\ \hline	
	MSDCR-w/o-DSAPE    & 0.1928          & 0.0927          & 0.2715          & 0.2574          & 0.1594          & 0.2989              & 0.2558          & 0.1668          & 0.3385          & 0.4571          & 0.3750          & 0.4238              \\
	MSDCR-w/o-DIAPE    & 0.2156          & 0.2110          & 0.2061          & 0.2734          & 0.2509          & 0.2450               & 0.2743          & 0.3124          & 0.2956          & 0.4555          & 0.4350          & 0.4146              \\
	MSDCR-w/o-Sep  & 0.2417          & 0.1291          & 0.1911          & 0.2955        &0.1870          & 0.2300           & 0.3293          & 0.2113          & 0.2833          & 0.4951          & 0.3904          & 0.4023             \\
	MSDCR-w/o-Enhan  & 0.1798          & 0.1340          & 0.1712          & 0.2454          & 0.1968          & 0.2194         & 0.2733          & 0.2019          & 0.2565          & 0.4750          & 0.3972          & 0.4039          \\
	MSDCR-Single-Target & 0.2748          & 0.2280          & 0.1509          & 0.3221          & 0.2723          & 0.2091          & 0.3581          & 0.2981          & 0.2174          & 0.5037          & 0.4348          & 0.3983           \\ \hline
	MSDCR        & \textbf{0.2967} & \textbf{0.2897} & \textbf{0.2965} & \textbf{0.3422} & \textbf{0.3305} & \textbf{0.3578} & \textbf{0.3711} & \textbf{0.3552} & \textbf{0.3483} & \textbf{0.5219} & \textbf{0.4821} & \textbf{0.4652}  \\ \bottomrule
\end{tabular}
	\caption{Performance comparison in Movie-Book-Music}
\label{Tbl:mov-book-msc}
\end{table*}

\begin{figure*}[t]
	\centering
	\subfigure[Movie-Music] {
		\begin{minipage}[t]{0.21\linewidth}
			\centering
			\includegraphics[scale=0.38]{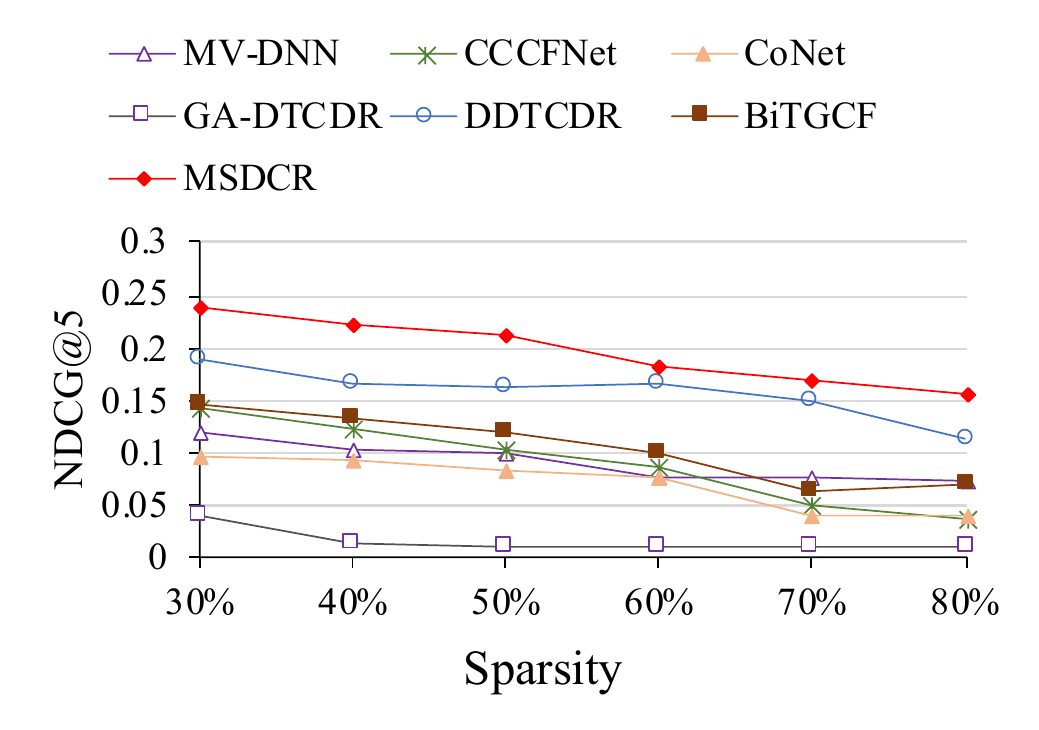}
			\label{fig:mov-msc}
		\end{minipage}%
	}
	\subfigure[Movie-Book] {
		\begin{minipage}[t]{0.21\linewidth}
			\centering
			\includegraphics[scale=0.38]{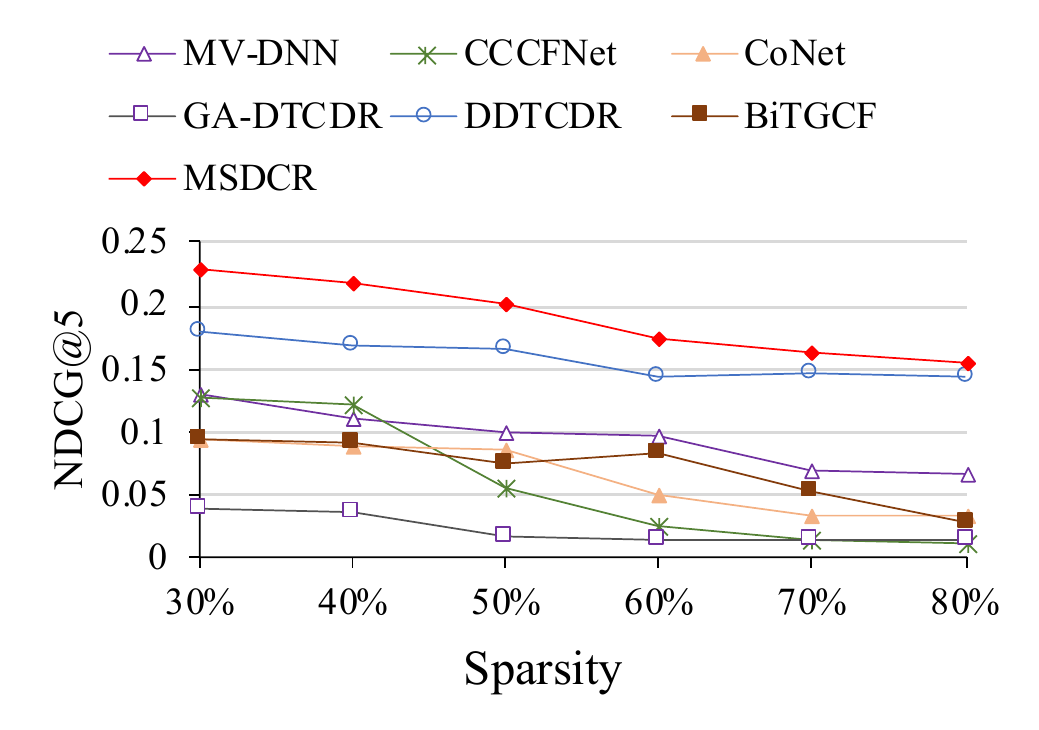}
			\label{fig:mov-book}
		\end{minipage}%
	}%
	\subfigure[Book-Music] {
		\begin{minipage}[t]{0.21\linewidth}
			\centering
			\includegraphics[scale=0.38]{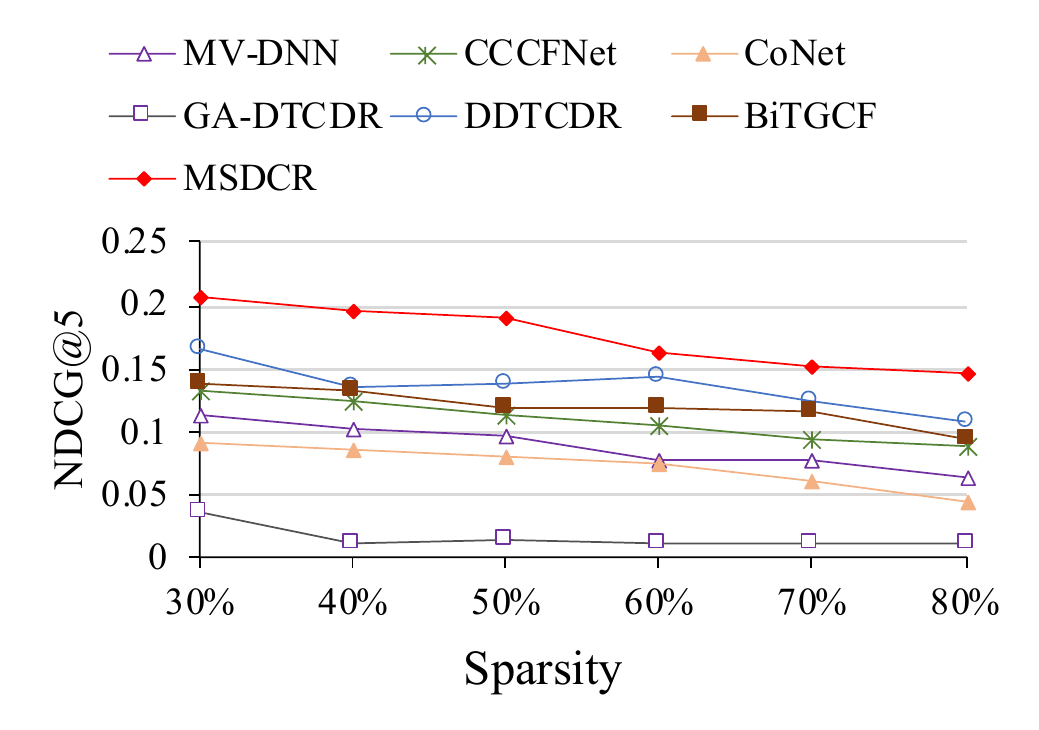}
	                 \label{fig:book-msc}
                  \end{minipage}%
	}%
	\subfigure[Movie-Book-Music] {
		\begin{minipage}[t]{0.21\linewidth}
%			\centering
			\includegraphics[scale=0.38]{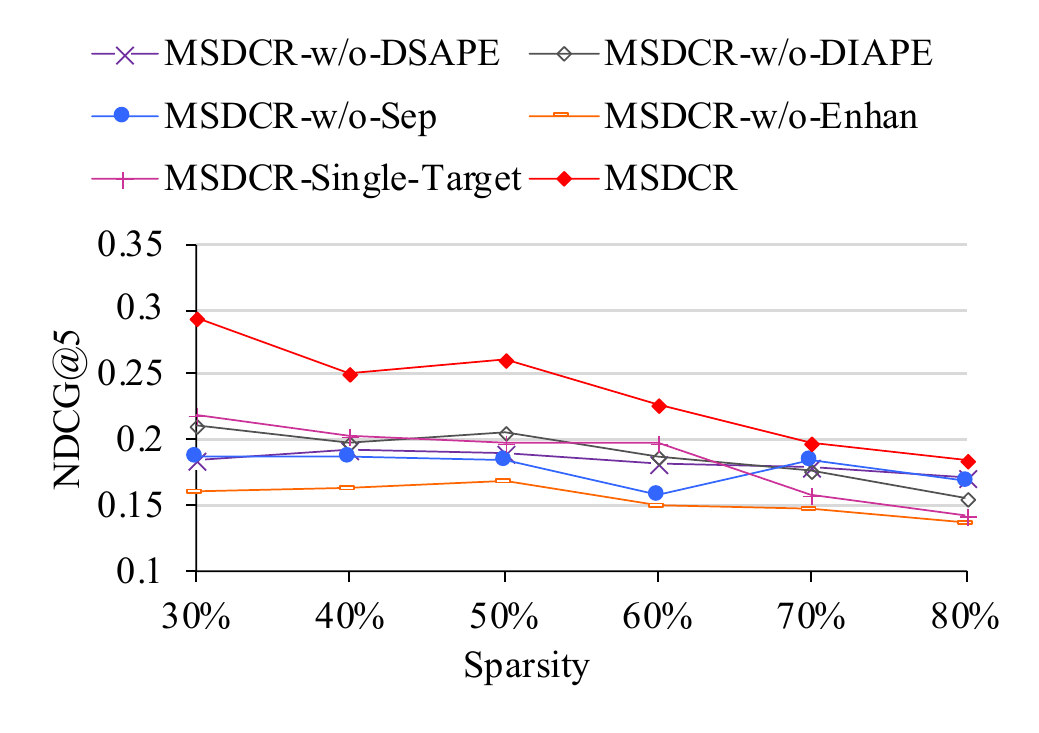}
	                 \label{fig:mov-book-msc}
       		 \end{minipage}%
	}%
	\caption{ Performance against sparsity.}
	\label{Fig:Sparsity}
\end{figure*}

\subsection{Ablation Study}
We use the triple-domain scenario Movie-Book-Music to check the effectiveness of the components of MSDCR by comparing it with its five variants: MSDCR-w/o-DIAPE (the variant without DIAPE), MSDCR-w/o-DSAPE (the variant without DSAPE), MSDCR-w/o-Sep (the variant where the domain separation is removed from DSAPE), MSDCR-w/o-Enhan (the variant where the gated preference enhancement is removed from DSAPE), and MSDCR-Single-Target (the variant where the multi-task framework is removed and MSDCR is trained separately for each single target domain). 

The results are shown in Table \ref{Tbl:mov-book-msc}. First, it can be observed that MSDCR performs remarkably better than MSDCR-w/o-DIAPE and MSDCR-w/o-DSAPE in all domains, which shows that a user's domain-specific preference and domain-invariant preference both play an importance role for the learning of a user's comprehensive preference in each domain. Second, we can see that MSDCR also outperforms MSDCR-w/o-Sep and MSDCR-w/o-Enhan in each domain. This verifies that the complementary aspect preference transfer via the gated aspect preference enhancement, together with the preserving of the uniqueness via the domain separation, can significantly improve the recommendation performance in each sparse domain. Third, we can note that MSDCR outperforms MSDCR-Single-Target, which shows the multi-domain collaboration is of benefit to MSDCR in multi-target CDR. At last, it is worthy note that for each domain, MSDCR performs better in triple-domain scenario (Table \ref{Tbl:mov-book-msc}) than in all dual-domain scenarios (Tables \ref{Tbl:mov-msc} - \ref{Tbl:book-msc}). This result verifies the superiority of MSDCR that the more sparse domains considered, the better the domains strengthen each other.

\subsection{Performance Against Sparsity}

Figures \ref{fig:mov-msc} - \ref{fig:mov-book-msc} show the average performances of MSDCR, the baseline methods and the variants in the three dual-domain scenarios and the triple-domain scenario, respectively, where the horizontal axes represent the proportion of the data that are randomly removed from the training set. We can observe that the performances of all methods degrade as the sparsity level increases (less training data). However, at most sparsity levels, MSDCR consistently performs better than all baseline methods, and the higher sparsity level, the bigger the performance gap between MSDCR and the baseline methods. The results demonstrate the superiority of MSDCR in the CDR scenarios where domains are all sparse, which is brought by its ability to build the comprehensive view of a user's preferences in each domain by transferring complementary preferences between the multiple domains.

\section{Related Works}
Single-target CDR aims at improving the recommendation performance in a sparse target domain with the help of a richer source domain.
Early methods for single-target CDR jointly factorize the rating matrices in relevant domains to generate the representations that can capture the common preferences of shared users \cite{singh2008relational,li2009can,hu2013personalized,lian2017cccfnet}. Recently, many DNN based methods have been proposed for better preference capturing across domains \cite{elkahky2015multi,man2017cross,hu2018conet,yuan2019darec,fu2019deeply,li2020ddtcdr,zhao2020catn,yu2020semi,du2020marline,gao2021cross,jin2021heterogeneous}, which often apply transfer learning techniques
like domain adaptation \cite{duan2009domain,sun2015survey,yu2020semi} to transfer domain-invariant preferences from a source domain to a target domain. 

In contrast to single-target CDR, dual-target CDR pursues better performance simultaneously on both two relevant domains \cite{zhu2019dtcdr}. The existing methods for dual-target CDR often apply bidirectional transfer learning to jointly model a user's preferences in two relevant domains \cite{hu2018conet,ma2019pi,zhu2019dtcdr,li2020ddtcdr,zhang2021model}. Recently, researchers have also proposed a few graph-based methods for CDR, which apply GNN to capture the high-order relations between features of different domains \cite{zhao2019cross,zhugraphical,liu2020cross}. PPGN \cite{zhao2019cross} leverages GCN to capture the high-order information propagation over the joint user-item interaction graph across different domains. GA-DTCDR \cite{zhugraphical} models the complex relations between users and items with a heterogeneous graph in each domain, and applies an element-wise graph attention network to fuse the embeddings of common users learned from both domains. BiTGCF \cite{liu2020cross} captures the high-order connectivity in user-item graph of single domain through GCN, and realizes a two-way transfer between two domains with the common user as anchors.

Basically, our MSCDR differs from the existing CDR methods in two aspects. First, the existing CDR methods, whether for single-target or for dual-target, often assume at least one domain has richer data, which might be impractical since sparsity is a ubiquitous problem. In contrast, our MSDCR treats the multiple domains as all sparse and can simultaneously improve recommendation performance for each domain. Second, the existing CDR methods heavily depend on the domain-invariant preferences, which might degrade the preference learning for multiple sparse domains. Different from the existing methods, MSCDR can build a comprehensive understanding of the user's preferences in each domain by considering both the domain-invariant preference and the enhanced domain-specific preference which are learned separately.

\section{Conclusion}
In this paper, we propose a Multi-Sparse-Domain Collaborative Recommendation (MSDCR) model for cross-domain recommendation. Unlike traditional CDR methods, MSDCR treats multiple relevant domains as all sparse and can simultaneously improve recommendation performance in each domain via comprehensive aspect preference learning. In particular, we propose a \textit{Multi-Domain Separation Network} (MDSN) and a \textit{Gated Aspect Preference Enhancement} (GAPE) module, which enables MSDCR to learn a user's unique enhanced domain-specific aspect preferences by adaptively transferring the complementary aspect preferences between domains. We also propose a \textit{Multi-Domain Adaptation Network} (MDAN) for MSDCR to capture the domain-invariant aspect preferences, which together with the enhanced domain-specific aspect preferences form a comprehensive view of a user's preference in each domain. At last, remarkable performance improvements on dual-target and multi-target CDR demonstrate the effectiveness of our MSDCR.

\begin{acks}
This work is supported by National Natural Science Foundation of China under grant 61972270, and NSF under grants III-1763325, III-1909323,  III-2106758, and SaTC-1930941.
\end{acks}

\newpage
\balance
\bibliographystyle{ACM-Reference-Format}
\bibliography{MSDCR}


\end{document}